\documentclass[runningheads]{llncs}

\usepackage{graphicx}
\usepackage{booktabs} 
\usepackage{enumitem}
\usepackage{xcolor,colortbl}
\usepackage{longtable}

\begin{document}

\title{A Survey on Software Engineering Practices in Brazilian Startups}

\author{Renata Souza\inst{1,2} \and
Orges Cico\inst{3} \and
Ivan Machado\inst{1}}

\authorrunning{Souza et al.}

\institute{Institute of Computing, Federal University of Bahia (UFBA), Salvador, Brazil
\\\email{\{renatamss, ivan.machado\}@ufba.br}\\
\and
Federal Institute of Bahia (IFBA), Santo Antônio de Jesus, Brazil
\email{\{renata.souza\}@ifba.edu.br}
\and
Norwegian University of Science and Technology (NTNU), Trondheim, Norway\\
\email{\{orges.cico\}@ntnu.no}}

\maketitle              

\begin{abstract}
Today's significant technological advancement allows early-stage software startups to build and launch innovative products quickly on the market. However, many of them die in the early years of their path due to market conditions, ignorance of customer needs, lack of resources, or focus, such as the misuse of well-established practices. The study's motivation is to analyze software engineering practices in startups from a practitioner's perspective. Our objective was to identify practices and tools the startups employ in their daily routines. We carried out an expert survey study with 140 software developers involved in software startups from different domains. The results show that startups in the initial and validation phases select practices and tools on an ad-hoc basis and based on the development team's prior knowledge. When they move into the growth phase, they recognize that they could have adopted better practices beforehand to support product scaling with a more mature team. The results also indicated that support tools are selected based on their integration with other tools and their ability to automate operational activities.

\end{abstract}

\section{Introduction} 
\label{section:introduction}

An impressive number of entrepreneurs create novel startups around the world every day. Later, they might become famous and profitable, such as \textit{Facebook, Instagram, Spotify, Linked-in, and Dropbox}. According to Sutton \cite{sutton2000role}, a startup shares many features with small and medium-sized young companies. For example, they are usually immature, have limited resources, undergo multiple influences, use dynamic technology, and hunt at the market. A startup can be defined as an organization established to search for a repeatable and scalable business model~\cite{blank2010startup}. Also, a startup is an organization dedicated to creating something new under conditions of extreme uncertainty~\cite{Ries2011lean}.

As they have limited financial and human resources, they face various challenges, and barriers \cite{Giardino2015,Klotins2015}. Most of them die in earlier years \cite{Ries2011lean}. Several critical challenges in every startup phase affect the strength and company stability \cite{Nguyen-DucWA2018}. It is crucial to prepare for what is to come so that an obstacle does not become unbeatable. Learning from others startups' experiences helps them find a way to succeed or fail faster \cite{Cavalcante2018,Crowne2002}. Therefore, it is necessary to leverage startup practices to understand the reason for these choices and their consequences.

This paper aims to shed light on the developers' perceptions of software engineering practices and tools employed by software startups in a Brazilian cohort. We carried out an expert survey study with 140 software developers involved in software startups from different domains. We sent an online questionnaire to professionals working in startups via email and social networks (\textit{Facebook, Twitter, and LinkedIn}). The results show that the software startups are guided by the customer, who participates from creating the initial prototype up to product deployment, and the adopted practices and tools that speed up development activities.

The remainder of this paper is structured as follows. Section \ref{section:background} covers the research-related background. Section \ref{section:methodology} introduces the survey study methodology. The results and the discussion are located in Sections \ref{section:results} and \ref{section:discussion}, respectively. Section \ref{section:implications} presents the implications for research and practice. ~The threats to this study's validity are discussed in Section \ref{section:threats}. Finally, Section \ref{section:conclusion} outlines overall conclusions and directions for future work.

\section{Background} 
\label{section:background}

\subsection{Startups}
Startups are companies that start from an early stage with very few assets, and low investment \cite{Blank2020,Giardino2016,Giardino2014software}. Some of them have the potential to achieve high scalability \cite{Unterkalmsteiner2016}. Many of these companies took advantage 
from the popularization of technologies for personal use, such as \textit{Google, Netflix, and Spotify}. They start almost from scratch \cite{Giardino2014framework,SouzaMA2017,SouzaRM2019sbqs} and could reach a wide range of customers ignored by major competitors offering solutions (almost always services) based on technology and the Internet \cite{Chanin2017}.

The startup concept definition, although simple, encompasses many variables. The primary way to recognize a startup refers to its business model characterized by an innovative idea that generates value and turns work into money \cite{Dash2019}. Additionally, it could be considered a policy aimed at developing competent, self-managed teams that could quickly react to scenario changes. The promotion of creativity, collaboration, focus on agility, continuous training, rewards for solving complex problems, and all actions must always link to a long-term vision.

\subsection{Startup lifecycle}

Blank \cite{Blank2020} defined the startup lifecycle as having four stages from a customer's perspective: (1) defining or observing a problem; (2) evaluating the problem; (3) defining a solution; and (4) evaluating the solution. Crowne \cite{Crowne2002} identified the startup lifecycle from the perspective of product development as having three stages, as follows:

\begin{enumerate}
    \item The \textit{startup} phase is the period between product conception and the first scale.
    \item \textit{Stabilization} begins when the first customer receives the product.
    \item \textit{Growth} is when the Startup delivers the product to new customers without creating any overhead in the development team.
    \item \textit{Maturity} is when the company evolves into a mature organization.
\end{enumerate}

Similarly, Wang \cite{wang2016key} defines a lifecycle for software startups taking into account the product stages: concept, underdevelopment, functional prototype, functional product with limited users, functional product with high growth potential, and mature product.

\subsection{Software Startups}

Software startups share some characteristics with other types of startups, such as lack of resources, operating history, high-technology challenges, cutting-edge tools, and innovation-driven development \cite{Unterkalmsteiner2016}. Carmel \cite{carmel1994time} was the first one to refer to the term \textit{software package startup} and the shorten term \textit{software startup} in 1994. Sutton \cite{sutton2000role} argues that software startups are challenged by limited resources, immaturity, multiple influences, high-technology, and turbulent markets. Hilmola \cite{hilmola2003value} states that most of the software startups are product-oriented and develop cutting-edge software products. Coleman \cite{coleman2008investigation} characterize software startups as product-driven, with a small development team, often developer-led, that develops software through various processes and without a methodology. Giardino \cite{Giardino2016}, and Unterkalmsteiner \cite{Unterkalmsteiner2016} use the term software startups to refer to those organizations focused on the creation of high-tech and innovative products, with little or no operating history, aiming to grow their business in highly-scalable markets aggressively. 

Large software companies relied for years on traditional software development. Nowadays, they realize they need to innovate to maintain their position in a highly competitive market. Also, they look to gain a significant advantage to adapt to market and technologies by focusing on cost efficiency, lead time reduction, and quality improvement. They find in startups this new way to keep their leading positions in a fast-moving and innovative market \cite{edison2018lean}. Software startups have become one of the critical drivers of economy and innovation, even though they are mostly inexperienced \cite{sutton2000role}. 

Coleman and O'Connor \cite{coleman2008investigation} investigated how software product startups establish their software development process. The study employed a grounded theory approach with the aim to characterized the experiences of small software organizations in developing processes to support their software development activities and the significant factors that might influence building processes: background of software, development manager background of founder, management style, process tailoring, market requirements. Hence, software startups depend mainly on the person's experience acting as software development manager who holds both expertise and know-how to meet company goals. Coleman and O'Connor \cite{coleman2008investigation} also claimed that startups are product-driven and developer-led. In this context, agile methods may have a lot to offer.

\subsection{Related Work}

Giardino et al. \cite{Giardino2015} and Wang et al. \cite{wang2016key} performed a large-scale survey of 5389 responses and an in-depth multiple-case study to investigate the critical challenges of software startups. Giardino et al. \cite{Giardino2015} observed the challenges in an early-stage startup, while Wang et al. \cite{wang2016key} observed challenges faced by software startups across lifecycle stages. Pantiuchina et al. \cite{Pantiuchina2017} also conducted an extensive survey of 1526 software startups where they examined the use of five agile and lean startup practices. In this work, we observed the software engineering practices adopted by startups through the divergent lifecycle stages. The startup's life cycle classification was based on the responses provided by the developers. This classification was made through characterization made concerning the information of the team size, the number of customers, and product maturity, shown in the following sections.

\section{Research Methodology} 
\label{section:methodology}

The study's motivation is to analyze software engineering practices in startups from a software practitioner's perspective. We chose a survey as our research instrument to reach a more significant number of opinions. Our goal in designing the study was to keep it as brief as possible while still gathering all relevant information. The survey received 140 responses. We did not ask for the identification of respondents or companies. We decided to preserve this information for reasons of confidentiality and attract more participants to our research. These participants were from all over the Brazilian territory. We applied the methodology recommended by Lin{\aa}ker et al. \cite{Linaaker2015} and the research survey principles defined by Kitchenham and Pfleeger \cite{Pfleeger2001}. This section encompasses the planning details, execution procedures, and reporting of the desired and achieved results.

\begin{figure*}[!ht]
\centering
\includegraphics[width=1.0\textwidth]{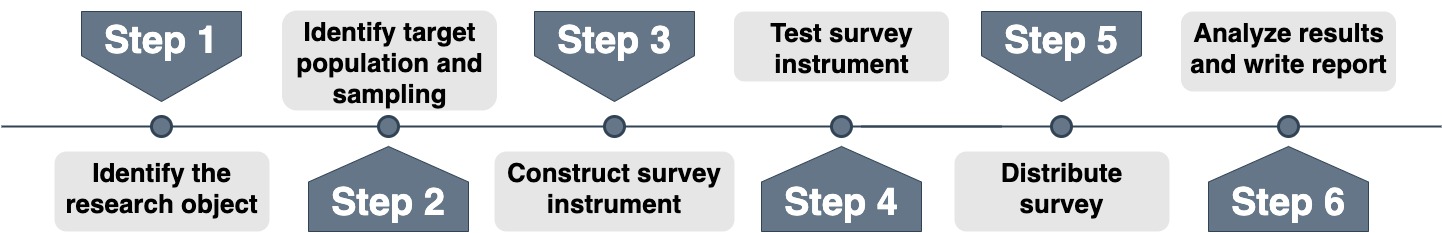}
\caption{Overview of the Research Methodology}
\label{figure:methodology}
\end{figure*}

We first identified the research object (Step 1), then the target population and sampling (Step 2). Next, we constructed the survey instrument (Step 3), tested it (Step 4), and distributed it to the participants (Step 5). The last action was to analyze the collected data (Step 6) and report the results. Figure \ref{figure:methodology} shows the research steps of this study. We utilized a web-based questionnaire to reach practitioners from Brazilian startup ecosystems due to Brazil's geographical dimensions.

\subsection{Identifying the Research Object} 
\label{subsection:questions}

Our objective was to identify software engineering practices and tools the startups employ in their daily routines. We also aimed to investigate whether there is any difference in practices and tools that depend on the startup's lifecycle. We formulated the following research questions:

\begin{enumerate}[label=\bf RQ\arabic*: ,leftmargin=1.4cm]
    \item \textbf{Do software startup use the same software engineering practices across different lifecycle stages?}
    \textit{We aim to leverage the practices software startups adopt through the different stages of their lifecycle and at what stage it differs.}
    \item \textbf{What tools do software startups use to support software engineering practices across different lifecycle stages?} 
    \textit{We aim to observe whether there is a change in the tool's adoption through the different stages of their lifecycle and at what stage it differs.}
\end{enumerate}

Startups need to either follow or adapt existing software engineering practices. Otherwise, they would risk losing their market capitalization. In this effect, further studies are needed to investigate both practices and tool support to establish a common understanding of how software startups make their choices.

\subsection{Identifying Target Population and Sampling}
\label{subsection:sampling}

The StartupBase\footnote{\url{https://startupbase.com.br}} is an official database of the Brazilian startup ecosystem supported by the Brazilian Startup Association\footnote{\url{https://abstartups.com.br/}}. Our target audience was software developers from software startups, while the population selected was Brazilian software startups. We chose software developers because they are the most representative group considering the respondents work for startups at different lifecycle stages and observe the same perspective. The survey included questions to reveal the participants' jobs, software development experience, and information on the organizations they worked for, such as size (ecosystem location) and project domain, to characterize the survey participants. We defined the following criteria to include participants in this survey study: (1) the respondent must work for a Brazilian software startup, and (2) the respondent must work as a software developer.

\subsection{Survey Instrument}
\label{subsection:instrument}

The questionnaire comprises closed questions with a small number of open questions to obtain further information. We specified twelve topics divided into two groups. The first was to gather the participation consent, respondents' characterization, startups' characterization, and product characterization. The second group covered the practices considering each software engineering discipline: software requirements, software architecture, software design, software construction, software testing, software maintenance, management, and processes. We next specify the goals defined for each category:

\begin{itemize}
    \item \textbf{Research participation consent:} free, prior and informed consent;
    \item \textbf{Respondents' characterization:} gather information about the respondents' profiles;
    \item \textbf{Startups' characterization:} gather information about when the startup was established, the number of customers, the revenue, the number of employees, and the business domain;
    \item \textbf{Product characterization:} characterize the product by type, public, lifecycle, revenue, need to pivot, and time that the team works on it;
    \item \textbf{Software requirements:} reveal the employed requirements engineering practices;
    \item \textbf{Software architecture:} identify decisions regarding system architecture;
    \item \textbf{Software design:} identify the decisions regarding user interface design;
    \item \textbf{Software construction:} leverage the main programming languages and development platforms the startup use;
    \item \textbf{Software testing:} leverage the startups' software testing practices;
    \item \textbf{Software maintenance:} leverage the employed maintenance techniques;
    \item \textbf{Project management:} leverage the project management practices;
    \item \textbf{Software process:} leverage the software process practices.
\end{itemize}

We used Survey Monkey\footnote{https://pt.surveymonkey.com/r/32FK9RP} to prepare and disseminate the questionnaire. Supplementary material is available at \cite{souza2021cibse}.

\subsection{Pilot Testing}
\label{subsection:testingsurvey}

We asked professionals and researchers with experience in software engineering and survey design to review the survey. The goal was to remove any misunderstanding and hence improve the instrument. The changes we made included: adding more answer choices to several questions, exchanging words to improve understanding, and changing some questions. We applied the questionnaire in September 2020. The questionnaire used at the end of the test contained the following questions:

\begin{enumerate}
\item Does the questionnaire contain anything expected to reach our goals?
\item Does the questionnaire contain any undesirable or unnecessary information to the context and purpose of the research?
\item Were you able to understand the questions properly?
\item Is there an error or inconsistency in the questionnaire?
\end{enumerate}

\subsection{Data Collection}
\label{subsection:distributing}

On September 2020, we sent the survey invitation by email, \textit{Facebook}, \textit{Twitter}, and \textit{Linked-in}. The invitation text contains the main instructions and the questionnaire link. We also sent reminders. We closed the survey in November 2020. We made a brief introduction with necessary information about the purpose of the study, justification of choice, and the importance of the respondent's participation. Participants were also informed about the privacy policies of the study in a clear and detailed manner.

\subsection{Data Analysis}
\label{subsection:analizing}

After assembling the data collection from the online survey, we took the following steps: \textit{(1) data validation} - this is the phase in which we analyzed the results to verify if the answers met the criteria of consistency and integrity; \textit{(2) participants' characterization} - data assembling and synthesis, including data they provided about the startups they work for as well as the software products or services their companies deliver; \textit{(3) data partitioning} - this is the phase in which the researchers analyze the responses and group them into subgroups (data partitioning counts on data obtained in the demographic questions, such as the number of customers, number of employees, the startups' diversification of software products or services, and the financial collection); and \textit{(4) data interpretation} - in this phase the quantitative analyses takes place \cite{Kitchenham2002}.

\section{Results} 
\label{section:results}

In this section, we report the results of our survey study. Some facts are worth mentioning from this survey application: all questions were mandatory, and we only analyzed responses whose respondents answered 100\% of the questions. One hundred fifty-eight people answered the questionnaire. However, 140 responses met our quality criteria.

\subsection{Characterization}
\label{section:Characterization}

\textbf{Respondents.} This section describes demographic data. We investigated the developers' education levels, expertise, roles, and working time to draw the observed sample profile. In terms of their professional experience in software development, 12\% had up to 1 year of experience, 31\% had between 2 and 3 years, 19\% had between 4 and 5 years, 15\% had between 6 and 10 years, and 23\% had more than 10 years of experience. Software developers classify themselves as related to their expertise; 27\% junior, 31\% mid-level, and 42\% were senior. Finally, the respondents played many different roles while working in startups: 58\% played more than one role against 42\%, who played only one role. 
{
}

\textbf{Product.} We highlight startups' products or services. 65\% startups characterized their products as Software-as-a-Service (SaaS), 31\% as a software product, and 4\% as a software and hardware product. 50\% of the startups developed only 1 product, and the other half had two or more products. 50\% startups developed a \textit{Business-to-Business} (B2B) product, 23\% a \textit{Business-to-Consumer} (B2C) product, and 19\% a \textit{Business-to-Business-to-Consumer} (B2B2C) product. 4\% claimed their product was both B2B and B2C. Finally, 4\% did not know or preferred not to answer. 
{
}

\textbf{Startups.} We also asked the respondents to provide us with information about the startups, such as the founding date, age, number of customers, revenue, number of employees, and lifecycle phases. 39\% of 26 startups have 1 to 5 employees; 17\% have 6 to 10 employees; 30\% have between 11 and 50 employees, and 13\% has more than 100 employees. Startups are from 13 different business domains. 57\% have between 1 and 10 customers; 22\% have between 11 and 100 customers; 9\% is between 101 to 100 customers, and 13\% has more than 1,000 customers. Regarding the startup lifecycle, 52\% of startups are in the early stage, 17\% of them in Validation, and 30\% in Growth. 
{
}

One of the size classifications of companies by revenue that the Brazilian Institute of Geography and Statistics (IBGE)\footnote{IBGE is the Brazilian agency responsible for official collection of statistical, geographic, cartographic, geodetic and environmental information in Brazil. Further information available at \url{https://www.ibge.gov.br/en/home-eng.html}.} uses organizes companies as follows: micro-entrepreneur, less than or equal to BRL 81K (USD 15.4K\footnote{Currency conversion on July 9th, 2021: BRL 1 = USD 0.19}); micro-enterprise, less than or equal to BRL 360K (USD 68.44K); small company, greater than BRL 360K and less than or equal to BRL 4.8M (USD 912.56K); medium-sized company, greater than BRL 4.8M and less than or equal to BRL 300M (USD 570M); large company, greater than BRL 300M.

\subsection{Startup Software Engineering Practices (RQ1)} 
\label{section:respondingsrqs}

To answer the research question, \textit{do software startup use the same software engineering practices across different lifecycle stages?}, we examined the occurrence of software engineering practices reported by software developers. 

\textbf{Software Processes.} We asked the professionals if they used to employ any formalized software engineering process. Most of them stated they commonly followed software development processes (61.5\%) and process management (50\%). 

\textbf{Software Methodologies.} Regarding the employed methodologies, we could observe that 96\% of developers perceive the agile methods as the best characterized their practices, followed by the iterative approach (23\%). Regarding the project planning activities they commonly used, the main reported ones were estimated at 70\%, followed by delivery planning with 65\%, scheduling with 54\%, resource allocation with 46\%, and process planning with 35\%.

\textbf{Requirements Engineering.} In every startup, we gathered data from all professionals reported any requirements engineering phase in their software development process. Nevertheless, only 28\% out of them stated their company had a formalized requirements engineering process. Requirements elicitation commonly employs prototyping in quick meetings with customers. In 80\% of the reports, the participants stated they typically involve customers in such elicitation tasks. In addition, 73\% said they used the brainstorming technique, 70\% of the participants said they used prototyping, among others.

\textbf{Design and Quality Attributes.} We asked the participants about software architecture concerns. The responses indicate that the two most used architectural patterns are model-view-control (46\%) and client-server (31\%). The participants mentioned that they used the layered architecture (27\%), ruled-based (15\%), and event-based (11\%) patterns. Other noted architectural patterns were piper and filter (8\%), reflection (8\%), peer-to-peer (4\%), broker (4\%), and batch (4\%). In terms of quality attributes (QA), 73\% of the participants identified usability as the most relevant QA. 65\% indicated that performance is a relevant QA as well. Additionally, 61\% pointed out that security/integrity and legal aspects are also relevant.

\textbf{Implementation.} Regarding approaches for developing/reusing source code, startups' software developers reported using frameworks (38\%) to speed up product/service development and delivery. They also claimed to use architectural patterns (35\%) and design patterns (35\%) as reuse strategies. 27\% claimed they built their products based on service-oriented systems (27\%) and program libraries (23\%). Next, configurable vertical applications (15\%), ERP (Enterprise Resource Planning) systems (12\%), program generators (12\%), software product lines (8\%), and packaging of legacy systems (8\%). Model-driven engineering and third-party components were other options presented by the survey participants, but none reported such practices.

\textbf{Software testing.} Regarding software testing practices, 69\% answered that they performed automated tests. Furthermore, 62\% said they use manual testing. Only 8\% responded that they did not perform any testing practices. When asked who runs the tests, in 69\% of the cases, the developers run the tests themselves. Conversely, only in 15\% of the cases was the developer who performed the test a person who did not implement the source code. Furthermore, 23\% claimed that customers ran tests. We also asked the developers about the test level they commonly performed. 77\% stated that their startups performed unit testing, 42\% performed component testing, 58\% performed integration testing, and 27\% performed acceptance testing. 8\% did not use to perform testing activities.

\textbf{Deploy}. When considering the time to deliver software, 19\% released it between 1 to 2 days, 12\% stated they released a software version between 3 to 7 days, 15\% between 1 and 2 weeks, 19\% between 3 and 4 weeks, and 16\% at two months or more. When it comes to delivering the software to the customer, most participants affirmed that 15\% spend between 1 to 2 days; 12\% between 3 to 4 weeks, (38\%) within 1 to 2 weeks; and 19\% percent reported that they released between 2 to 3 months.

\textbf{Maintenance.} Regarding the code maintenance techniques adopted in the startups, 38\% stated that they did not use any software maintenance techniques. Among the startups that used these techniques, 38\% said they only used code understanding to maintain products or services, 27\% said they used migration, 15\% applied to re-engineer, 4\% used reverse engineering, and 4\% discontinued the application.

\subsection{Tool Support (RQ2)} 

To answer the research question, \textit{what tools do software startups use to support software engineering practices across different lifecycle stages??}, we examined the occurrence of tools reported by software developers. Table \ref{table:tools} shows the set of support tools pointed out by the participants. They are helpful to handle the following project lifecycle phases: software requirements, software architecture, software design, software development, software testing, software configuration management, and project management. The choice of tools that support software development seems to be adopted primarily according to the team's awareness of existing tools.

In addition, we could enlist a set of reasons why the participants decided to choose a particular support tool to use in their projects: the tool should allow (and improve) the communication between team members; allow rapid changes in the development and testing of the software product or service to meet customers' needs; allow prototyping of the product or service; automate development activities (such as automated versioning, code integration, continuous deployment, and automatic build generation); and allow the analysis of data generated by users to generate knowledge about their preferences.

\begin{table*}[!ht]
\centering
\caption{Support tools per software development  lifecycle phase.}
\def \arraystretch{1.1}
\label{table:tools}
\resizebox{.9\textwidth}{!}{
\begin{tabular}{c|c|c|c|c|c|c} \toprule    
     \rotatebox{90}{\bf \begin{tabular}[c]{@{}l@{}}Software\\ Requirements \end{tabular}} & 
     \rotatebox{90}{\bf \begin{tabular}[c]{@{}l@{}}Software\\ Architecture \end{tabular}} & 
     \rotatebox{90}{\bf \begin{tabular}[c]{@{}l@{}}UI Design \end{tabular}} & 
     \rotatebox{90}{\bf \begin{tabular}[c]{@{}l@{}}Software\\ Development \end{tabular}} & 
     \rotatebox{90}{\bf \begin{tabular}[c]{@{}l@{}}Software\\ Testing \end{tabular}} & 
     \rotatebox{90}{\bf \begin{tabular}[c]{@{}l@{}}Software\\ Configuration\\ Management \end{tabular}} & 
     \rotatebox{90}{\bf \begin{tabular}[c]{@{}l@{}}Project\\ Management \end{tabular}}\\ \midrule  
\begin{tabular}[t]{@{}l@{}}JIRA\\  DOORS\\ GitLab\\ MS Office \end{tabular} &
\begin{tabular}[t]{@{}l@{}}UML Designer\\ Papyrus\\ Diagramo\\ Plant UML\\ draw.io \end{tabular} &
\begin{tabular}[t]{@{}l@{}}Adobe XD\\ Sketch\\ Zepplin\\ Balsamiq\\ Pencil\\ Axure\\ Fluid\\ In Vision \end{tabular} &
\begin{tabular}[t]{@{}l@{}} TypeScript\\ Android Studio\\ Eclipse\\ JetBrains\\ NetBeans\\ 
                           PyCharm\\ Unity\\ RStudio\\ Visual Studio\\ XCode \end{tabular} & 
\begin{tabular}[t]{@{}l@{}}TestComplete\\ Selenium\\ JUnit \\ JIRA\\ TestFLO \end{tabular} &
\begin{tabular}[t]{@{}l@{}}GitLab\\ Bitbucket\\ Subversion\\ GIT \end{tabular} & 
\begin{tabular}[t]{@{}l@{}}JIRA\\ Trello\\ GitLab \end{tabular} \\  
\bottomrule
\end{tabular}}
\end{table*}

\section{Discussion} 
\label{section:discussion}


The survey results revealed that software professionals focus their efforts on speeding up the construction of the software product based on customer guidelines and concentrate mainly on usability, performance, and security. Project planning is primarily based on deliveries and validations with their customers, who become involved in software development activities, from eliciting requirements through prototypes to testing and delivering the product or service. Although architectural styles and design patterns are prevalent, software architecture decisions are based on the software development team's knowledge in an \textit{ad-hoc} manner. Software development is based on agile practices, regardless of other well-established software development processes. Also, the development process is mainly schedule-driven and focused on planning for delivery to the customer. When it comes to productivity, we found a general interest in employing frameworks and third-party API projects.


We investigate the differences between software practices adopted by software startups to their lifecycle stage.
We observe that startups adopt a similar set of practices and tools in the three stages: startup, validation, and growth. However, looking at the frequency of responses, we notice that they value more practices and tools that support seamless integration, automated testing, automated security verification, and faster delivery to the user as they enter the growth stage. Startups realize that choices made in earlier stages have affected software quality and software development speed. And now, a new challenge is to solve the accumulated problems while adopting good software development practices, minimal documentation, and automated testing to provide security and meet the growing needs or demands of the increasing customer base.

\section{Implications for Research and Practice} 
\label{section:implications}

In this section, we present the relevant implications that emerge from the analysis of this study:

\begin{itemize}
    \item \textbf{Research Implications.} The discoveries made by the academic and scientific community must be transferred to software startups. Software startups work in an innovative and complex environment and need recommendations on which best practices and tools to adopt at each stage of their lifecycle and recommendations on avoiding pitfalls that affect the short and long run in a direct, clear and precise way.
    \item \textbf{Practice Implications. } By observing how they choose and use software engineering, we identified a startup culture. The culture is how they see and act in which they permit themselves not to have the correct answers for every question and difficulty. They incessantly focus on the customer's pain and how to solve it but focus on finding the right question to be resolved. In this way, they have the opportunity to construct an answer or a solution that fits better that can be validated and refined with the customer's presence. There is the possibility of making mistakes on this path, starting over and trying again more assertively to build an answer or a solution with relevance. 
\end{itemize} 

{
}

\section{Threats to Validity} 
\label{section:threats}

\textit{Construct Validity}: During the pilot test, some respondents reported that the instrument's filling time was extensive. As such, our survey respondents may not have adequately answered questions, preferring short answers to more detailed descriptions. We grouped the questions into specific sections to better target questions and answers to reduce threats to validity. Another threat was the respondents' understanding of the questions. To help ensure the survey's understandability, we asked professionals and researchers with experience in software engineering and experience in survey design to review the study to ensure the questions were clear and complete.

\textit{Internal Validity}: Another observed threat may be selecting practitioners to the sample. We understand that the number of responses obtained may not adequately represent the entire population of startup software professionals, characterizing a threat to internal validity. However, as we decided to include only professionals from startups that work in different domains (and mostly have offices in several Brazilian cities), we believe this set might represent. 

\textit{External Validity}: The respondents of our survey may not adequately represent all startup software practitioners. Thus, our results could not be statistically relevant. Nevertheless, we believe that the 140 responses we analyzed provide a rich qualitative data source to reveal valuable insights.

\textit{Reliability}: This threat indicates that the interpretation might influence the research results. Two authors of this paper carried out the analysis process by working together to mitigate such a threat. We discussed any disagreements in the assignment until we reach a consensus.

\section{Conclusion} 
\label{section:conclusion}

Software startups are born with disruptive ideas and business models. Some of them have to pivot until they find a valuable and scalable business model. They need to find a strategy to scale and enter a growth phase decreeing its continuity or closure. However, there are significant challenges software startups commonly to reach higher levels on the growth scale. Among the challenges, improving software engineering practices is a demanding task for software startups.

This study analyzes how Brazilian software startups have dealt with software engineering practices from a software practitioner's perspective. We surveyed 140 responses from software developers working in Brazilian software startups. Our study shows that startups in the initial and validation phases choose software engineering practices on an \textit{ad-hoc} basis and primarily based on the development team's knowledge. When they move into the growth phase, they recognize that they could have adopted better software development practices to support the product scale with a more mature team. The developers decide to use support tools that integrate with others and automate operational activities. Future work directions should consider extending this study to reach professionals from other global regions to perform a cross-validation analysis to verify whether the results are still valid when considering different scenarios.

\subsubsection*{Acknowledgments.}
This research was partially funded by INES 2.0; CNPq grants 465614/2014-0 and 408356/2018-9 and FAPESB grant JCB0060/2016.

\bibliographystyle{splncs04}

\end{document}